\documentclass{aa}
\usepackage{graphicx}
\usepackage{natbib}
\usepackage{txfonts}

\newcommand{\ppm}{\mbox{$\pm$}}
\def\cts{counts~s$^{-1}$}
\def\x8{M33~X-8}

\begin{document}

\title{High resolution {\em Chandra} X-ray imaging of the nucleus of M33}

\author{G. Dubus
	\inst{1,2}
	\and
	P.~A. Charles
	\inst{3}
	\and
	K.~S. Long
	\inst{4}}

\institute{Laboratoire Leprince-Ringuet, Ecole Polytechnique, Palaiseau F-91128, France; \email{gd@poly.in2p3.fr}
\and
Institut d'Astrophysique de Paris, 98bis boulevard Arago, Paris F-75013, France
\and
Department of Physics and Astronomy, University of Southampton, Southampton SO17 1BJ, UK
\and
Space Telescope Science Institute, 3700 San Martin Dr., Baltimore MD 21218}

\date{Draft \today}
\abstract{
We show with a short {\em Chandra} HRC-S exposure that the
powerful X-ray source coincident with the nucleus of M33 is
unresolved at the highest spatial resolution available to date
(0\farcs4). The source flux is variable, doubling during the 5~ks
exposure. The combination of properties exhibited by \x8
establishes the source as the nearest example of the
ultra-luminous X-ray sources that have been uncovered in other
nearby galaxies.  On short timescales, we set limits of 9\% r.m.s.
variability for pulsations in the 0.01--1000~Hz range.

}
\maketitle

\section{Introduction}
The nearby galaxy M33 ($d\approx795$~kpc; \citealt{vdb}) hosts the
most luminous steady X-ray source in the Local Group: \x8 has a
1-10~keV luminosity of 1.2$\cdot 10^{39}$~erg~s$^{-1}$. The source has
been detected at this level ever since the first X-ray observations of
M33 \citep{long,markert}.

{\em Chandra} observations established that \x8 is coincident with the
optical position of the nucleus to the 0\farcs6 uncertainty of the
astrometric solution \citep{dr02}. However, the 1500~M$_\odot$ upper
limit on the mass of a central black hole placed by the measured
velocity dispersion \citep{gebhardt} and the lack of activity at other
wavelengths (\citealt{lcd} and references therein) rule out a low
luminosity active galactic nucleus.

Ultra-luminous X-ray sources (ULXs) are defined as point sources
having X-ray luminosities in excess of the Eddington limit for a 10
M$_\odot$ object ($10^{39}$~erg~s$^{-1}$; \citealt{fabbiano}).  ULXs
could be due to beamed emission from stellar mass-sized neutron stars
(NS) / black holes (BH) or truly isotropic emission from
100-1000~M$_\odot$ intermediate mass BHs \citep{king}. If \x8 is a
single object, then it is the nearest example of a ULX. The detection
of a 106 day $\sim$20\% modulation in the X-ray flux from \x8 in a 5
year span of {\it ROSAT} data \citep{106} and the resemblance of the
spectrum to that of black hole binaries \citep{laparola} support this
picture.

{\it ROSAT} images show \x8 to be point-like at 5\arcsec\
resolution \citep{schulman}. But diffuse emission from the compact
optical nucleus (3\arcsec\ wide, see Fig.~3 in \citealt{kormendy})
or contributions from close, faint sources would not have been
resolved using any pre-{\em Chandra} instrument. With its superior
angular resolution, {\em Chandra} has been able to reduce the
possibility that ULXs in nearby galaxies are chance superpositions
of lower luminosity sources.  Unfortunately, in the case of \x8,
the existing {\it Chandra} observations of \x8 had the nuclear
source either far off-axis or heavily piled-up in ACIS CCD. As a
result, the Chandra data could not be used to study the radial
extent below a few arc seconds.

Here we report a new observation of \x8 with {\em Chandra}, an on
axis observation obtained with the { High Resolution Camera} (HRC)
on board {\em Chandra} that fully exploits the spatial resolution
of {\em Chandra} in an attempt to resolve the nuclear source
in M33. The observation, spatial and timing analysis are described
in \S2-4. The results are discussed in \S5.

\section{Observation}

M33 was observed with the {\em Chandra}/HRC-S instrument \citep{hrc1} on
July 29, 2003 from 19:53:12 UT for a total exposure time of
5355~s. The HRC is unaffected by pile-up and provides the best
achievable spatial resolution (0\farcs4 FWHM) to date. We decided to
use the HRC-S which has the capability to study the poorly known
timing properties of \x8 at frequencies above 1~Hz.   We performed the analysis
with CIAO version
3.0.1\footnote{{\tt http://asc.harvard.edu/ciao}}.

There is only one point source, \x8, with signal-to-noise greater than
5 in the field-of-view. In the $6\times8$\arcmin\ central region
surrounding \x8 where the background is spatially uniform, we
estimated the average background rate per 2\arcsec\ circular region,
and found that 28 counts would be needed for a 3$\sigma$ detection
(taking into account the 13750 separate trials).  This translates into
a 0.6-9~keV luminosity upper limit for point source detection in this
observation of $< 9~10^{36} (d/795~{\rm kpc})^2$~erg~s$^{-1}$
(assuming a photon power-law of $\alpha$=2, $N_H=1.9\times 10^{21}
{\rm cm}^{-2}$). There are 43 sources within 8\arcmin\ of \x8 in the
X-ray {\em ROSAT} catalogue of \citet{haberl} but the brightest of
these (excluding X-5 which is at the very edge of the field-of-view)
have count rates a factor $\la$ 1\% that of \x8.  Hence, it is not
surprising that we failed to detect any other sources in this short
exposure.

\section{Timing Analysis}
\begin{figure}
\resizebox{\hsize}{!}{\includegraphics{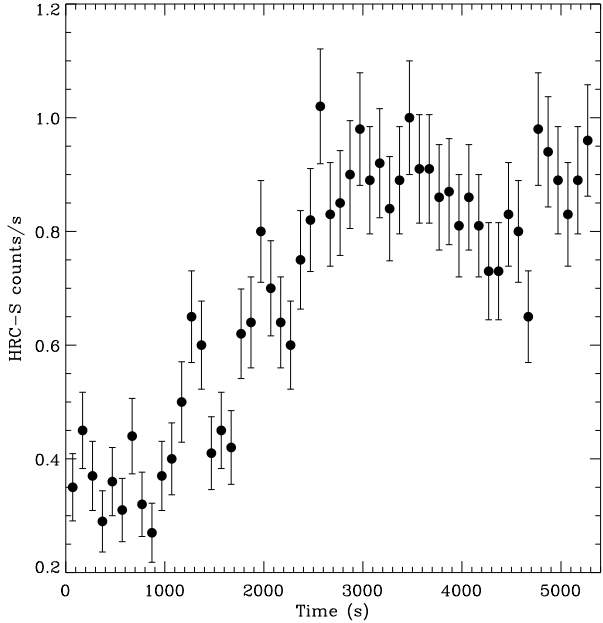}} 
\caption{HRC-S
lightcurve of \x8 in 100s bins, showing significant
variability on timescales of $\sim$ 1~ks.}
\label{fig:lc}
\end{figure}

We extracted 3756 counts from a 16 pixel radius region around \x8,
giving an average of 0.70\ppm0.01 \cts. Using an annulus with inner
and outer radii of 20 and 30 pixels, we estimate the background
contribution to be only 6 counts (and hence negligible).
Surprisingly, the count rate from \x8 varied significantly during our
observation, increasing from 0.36\ppm0.03 to 0.92\ppm0.04 \cts, on a
timescale of 2400~s (Fig.~\ref{fig:lc}). Previous results have
consistently shown that the X-ray spectrum of \x8 is well fitted by a
$kT$=3.5 keV thermal bremsstrahlung spectrum with
$N_H=1.9\times10^{21}~{\rm cm}^{-2}$
\citep{takano,parmar,dr02,laparola}. Assuming this spectrum (for which
1 HRC-S count corresponds to 2.6$\cdot 10^{-11}$~erg~cm$^{-2}$ of
unabsorbed flux in the 0.6-9 keV band), the maximum count rate is
equivalent to $L_X$ = 1.8$\cdot 10^{39}~(d/795~{\rm
kpc})^2$~erg~s$^{-1}$.

\begin{figure}
\resizebox{\hsize}{!}{\includegraphics{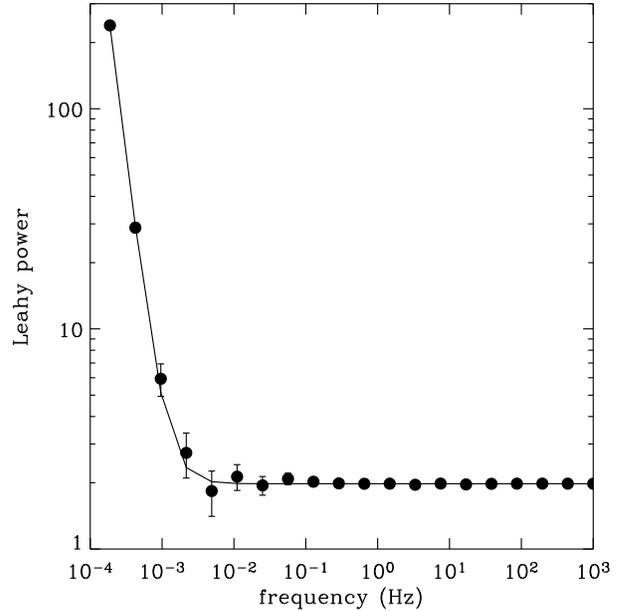}}
\caption{Power Density Spectrum of \x8\ during the 5355~s {\em
Chandra} HRC-S observation, normalized according to \citet{leahy83}.
Data have been rebinned logarithmically.  The solid line shows the
best-fit power-law (slope $\alpha$=2.65\ppm0.15) + constant noise
(taken equal to the average at high frequencies, 1.98).}
\label{fig:pds}
\end{figure}

We applied barycenter corrections to the photon arrival times and
computed their power density spectrum (PDS), with a Nyquist frequency
of 1000~Hz. We find no excess power above 0.01~Hz, with a 99.9\% upper
limit of $<$8.8\% root-mean-squared (r.m.s.)  variability from a
sinusoidal signal.  Logarithmically rebinning the PDS to 100 bins, we
do not find any excess from broad timing features above
$>$0.01~Hz. The 90\% upper limit to a quasi-periodic oscillation (QPO)
is 24\% r.m.s. where we have modeled the QPO as a 0.25~Hz FWHM
gaussian at 6~Hz. There is significant power at low frequencies with a
measured 30\ppm6\% r.m.s. variability below 0.01~Hz.  Further
rebinning to 20 frequency bins (Fig.~\ref{fig:pds}), we find that the
data are well described ($\chi^2_\nu$=0.7, 18 degrees of freedom) by a
fit to a constant plus a power-law low frequency noise (LFN) with a
steep slope $\alpha$=2.65\ppm0.15 (90\% confidence error). The
constant is taken equal to the high frequency average, 1.98 (close to
the expected level from a pure Poisson flux).  However, the LFN power
must turn over between 0.02 and 0.19~mHz in order for it not to
diverge (i.e.  exceed 100\%). To constrain any additional high
frequency, flatter timing feature, we fixed the power-law slope at
-2.65 and added a second component with $\alpha=1$. The 90\%
confidence upper limit to the r.m.s. variability below 10~Hz is
$<$3.2\%, corresponding to a change in the minimum $\chi^2$ of 4.61
\citep{lampton}.

In spite of the lack of photon energy sensitivity in the HRC
detectors, we can still search for spectral variability by exploiting
the fact that the photon energy to pulse-invariant (PI) energy channel
calibration is stable during an observation.  We therefore
investigated the data in two ways: firstly by comparing counts in the
first 2000~s interval (cf. Fig.~\ref{fig:lc}) with those detected
afterwards (a total of 907 and 2829 counts respectively); secondly by
dividing in time so as to produce an equal number of counts in each
group (which occurs 3172~s into the observation).  The distributions
of the PI channels between groups are statistically identical
(respectively a probability of 0.30 and 0.19 of being drawn from the
same distribution for a two-sided Kolmogorov-Smirnov test).  Hence,
within the constraints of the HRC, there is no evidence for spectral
variability during our observation.

\section{Spatial Analysis}

Our main purpose in undertaking this observation was to determine
whether the nuclear source in M33 is point-like.  We therefore used
{\tt ChaRT} and {\tt Marx} to simulate how the PSF of \x8 would appear
on the HRC-S detector if it were point-like.  In carrying out this
simulation (of 28861 counts in total), we assumed the same spectrum as
in \S3.  This simulation accounts for all known telescope and detector
effects.  It also shows why the spectrum is important, as the {\tt ChaRT} PSF is significantly broader than a monochromatic
PSF (at 1~keV) calculated with {\tt mkpsf}.

\begin{figure}
\resizebox{\hsize}{!}{\includegraphics{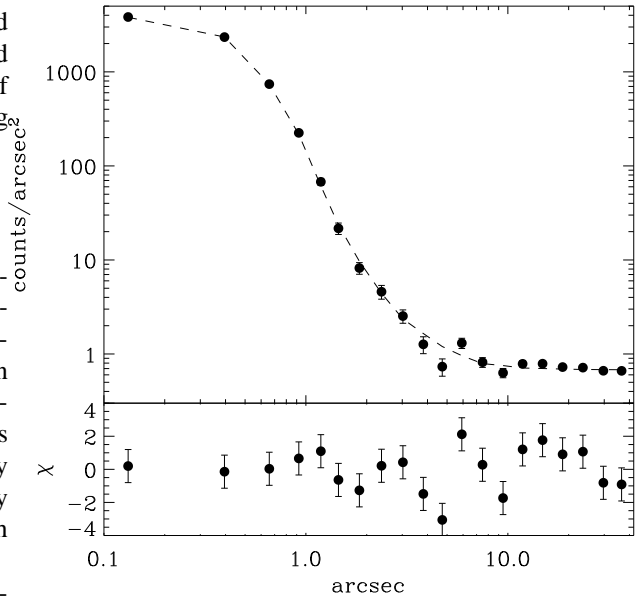}} 
\caption{{\bf Top Panel:} The observed radial profile of \x8, showing
the number of counts (with Gaussian errors) per unit area in annular
bins.  The dashed line is the best fit model (from {\tt ChaRT}), and
includes a constant background component.  \x8 dominates at radii
$<$4\arcsec, while the background dominates at larger radii.  {\bf
Bottom Panel:} The $\chi$=(data-model)/$\sigma$ residuals.  We
attribute the scatter between radii of $5$ and $40$\arcsec as due to
unresolved point sources.}
\label{fig:psf}
\end{figure}

We logarithmically re-binned the data into a 1-D radial distribution
(between $r=0$ and $40$\arcsec) for comparison with a {\tt ChaRT}
simulation (Fig.~\ref{fig:psf}).  We determined the source centre by
rebinning the PSF and performing a radial fit to the data, minimising
$\chi^2$ against the 2-D offset. The best-fit to the PSF model has
$\chi^2_\nu$=1.72 (18 degrees of freedom), and the simulated PSF is a good
description of the data. We find no evidence for any additional
component to the emission profile.

Extended emission associated with the nucleus is not statistically
required by the data. To place an upper limit on its contribution, we
convolved the PSF with the radial profile derived from {\em Hubble
Space Telescope} visible data \citep{dubus}. The 90\% confidence upper
limit is $<$3.7\% of the average flux from \x8 during the observation
or $4.7\cdot 10^{37}$~erg~s$^{-1}$.

We also examined the residuals obtained by subtracting the best-fit
PSF from the data using an image bin size of 3 pixels (0\farcs4),
equal to the resolution of the HRC-S. There is no apparent structure
in the residuals, which have amplitudes smaller than $<$7\% of the
peak flux. We derive an upper limit of $9.1\cdot 10^{37}$~erg~s$^{-1}$
to the contribution from a point source between 0\farcs4--1\arcsec\
from the nucleus.

\section{Discussion}

\cite{dr02} established to a high degree of precision that M33 X-8
is coincident with the nucleus.  The present {\em Chandra} HRC-S
observation shows that \x8 is also unresolved at 0\farcs4
resolution, improving on earlier {\em ROSAT} HRI upper limits by a
factor $\sim$ 10 \citep{schulman}. This rules out contributions
from nearby sources but does not rule out multiple sources within
the compact nucleus core \citep{dubus}. However, the substantial
(50\%) variability we observe during this short 5~ks observation
shows that a single source dominates. We conclude \x8 is most
likely a single source and, as has been claimed previously, the
nearest example of a ULX source.

A possible key to uncovering the nature of this source is its
variability. The short timescale doubling that we observe is
unprecedented: within single pointed observations \x8\ is usually
seen to be constant. Variability at a level of 10\% has been
observed within hours to days in only two {\em ROSAT} visits
\citep{106}. More recently, 10\% variability on a timescale of
5000~s was seen in a continuous 93~ks {\em Chandra} off-axis
observation (\citealt{laparola}; see also \citealt{peres} for an
earlier report). Although the timescale of the variation we
observe is similar, the amplitude in our observation is much
larger than the 10\% variation observed by \cite{laparola}.

Using the doubling timescale of 2400~s, we can set an upper limit
to the emission region of 7$\cdot10^{13}$~cm which hints at a
compact object. Recently, it has been proposed that the break
frequency at which the PDS of active galactic nuclei and X-ray
binaries switch from a -2 to a -1 power-law slope depends on the
black hole mass \citep{mchardy}. Using the fact that the PDS of
\x8 must turn over between 0.02 and 0.19~mHz, and interpreting
this break as the high frequency turnover ({\em e.g.} NGC 4051,
\citealt{mchardy}), we find the black hole mass would be in the
range $10^{5-6}$~M$_\odot$. This is inconsistent with velocity
dispersion upper limits \citep{gebhardt}. Hence, it is unlikely
the variability is associated with the high frequency noise phenomenon
described by \cite{mchardy}. Much longer exposures are needed to
establish the timing properties of \x8 and investigate its nature.

The X-ray spectrum, long timescale variability and lack of an optical
counterpart suggest an X-ray binary. The $\sim$1 hour doubling, but
lack of fast variability, are consistent with a low mass X-ray binary
in the high or very high state. On long timescales, \cite{markert}
observed a 50\% count rate drop between two {\em Einstein} HRI
observations separated by six months. Later {\em ROSAT} observations
confirmed variations of $\ga 20$\% on a timescale of 106~days
\citep{106}. \citet{laparola} have emphasized the similarity between
\x8 and LMC~X-3. LMC~X-3 has long term variations on a 99~day (or
199~day) timescale \citep{wilms}. We note that the present observation
would correspond to phase $\phi=0.87$ in Fig.~3 of \citet{106}, very
close to the minimum of the cycle where the folded {\em ROSAT} data
shows the most scatter.

{\em Chandra} results indicate that ULXs are more frequently found
in association with recent star formation \citep{fabbiano}. There
is evidence for a 40~Myr old starburst in the nucleus of M33
\citep{oconnell,lcd}. The presence of \x8 could be linked to this
episode of star formation, suggesting perhaps a source comprised
of a high mass companion orbiting a stellar-mass black hole. At
present, the source in the nucleus of M33 appears to have all of
the characteristics of a canonical ULX. As such, a better
understanding of the nearest ULX should remain a priority.

\begin{acknowledgements}
Support for this work was provided by NASA through
{\em Chandra} Award Number NAS8-39073 issued by the {\em Chandra} X-ray
Observatory Center, which is operated by the Smithsonian Astrophysical
Observatory for and on behalf of NASA under contract NAS8-39073.
\end{acknowledgements}


\begin{thebibliography}{}
\bibitem[Dubus \& Rutledge(2002)]{dr02}Dubus, G., \& Rutledge, R.E., 2002, \mnras, 336, 901
\bibitem[Dubus et al.(1999)]{dubus} Dubus, G., Long, K.~S., \& Charles, P.~A.\ 1999, \apjl, 519, L135
\bibitem[Dubus et al.(1997)]{106} Dubus, G., Charles, P.~A., Long, K.~S., \& Hakala, P.~J.\ 1997, \apjl, 490, L47
\bibitem[Fabbiano \& White(2003)]{fabbiano} Fabbiano, G., \& White, N.~E., 2003, to appear in {\em Compact Stellar X-ray Sources}, Eds. Lewin, W.~H.~G., \& van der Klis, M., Cambridge University Press (astro-ph/0307077)
\bibitem[Gebhardt et al.(2001)]{gebhardt}Gebhardt, K.~et al.\ 2001, AJ, 122, 2469
\bibitem[Haberl \& Pietsch(2001)]{haberl} Haberl, F.~\& Pietsch, W.\ 2001, \aap, 373, 438
\bibitem[King et al.(2001)]{king} King, A.~R., Davies, M.~B., Ward, M.~J., Fabbiano, G., \& Elvis, M.\ 2001, \apjl, 552, L109
\bibitem[Kormendy \& McClure(1993)]{kormendy} Kormendy, J.~\& McClure, R.~D.\ 1993, \aj, 105, 1793
\bibitem[La Parola et al.(2003)]{laparola} La Parola, V., Damiani, F., Fabbiano, G., \& Peres, G.\ 2003, \apj, 583, 758
\bibitem[Lampton, Margon, \& Bowyer (1976)]{lampton} Lampton, M., Margon, B., \& Bowyer, S. \ 1976, \apj, 208, 177
\bibitem[Leahy et al.(1983)]{leahy83} Leahy, D.~A., Darbro, W., Elsner, R.~F., Weisskopf, M.~C., Kahn, S., Sutherland, P.~G., \& Grindlay, J.~E.\ 1983, \apj, 266, 160
\bibitem[Long et al.(2002)]{lcd} Long, K.~S., Charles, P.~A., \& Dubus, G.\ 2002, \apj, 569, 204
\bibitem[Long, D'Odorico, Charles, \& Dopita(1981)]{long} Long, K.~S., D'Odorico, S., Charles, P.~A., \& Dopita, M.~A.\ 1981, \apjl, 246, L61
\bibitem[Markert \& Rallis(1983)]{markert} Markert, T.~H., \& Rallis, A.~D., 1983, \apj, 275, 571
\bibitem[McHardy et al.(2004)]{mchardy} McHardy, I.~M., Papadakis, I.~E., Uttley, P., Page, M.~J., \& Mason, K.~O.\ 2004, \mnras, 348, 783
\bibitem[Murray et al.(1998)]{hrc1} Murray, S.~S., Chappell, J.~H., Kenter, A.~T., Kraft, R.~P., Meehan, G.~R., \& Zombeck, M.~V.\ 1998, \procspie, 3356, 974
\bibitem[O'Connell(1983)]{oconnell} O'Connell, R.~W., 1983, \apj, 267, 80
\bibitem[Parmar et al.(2001)]{parmar} Parmar, A.~N.~et al.\ 2001, \aap, 368, 420
\bibitem[Peres, Reale, Collura, \& Fabbiano(1989)]{peres} Peres, G., Reale, F., Collura, A., \& Fabbiano, G.\ 1989, \apj, 336, 140
\bibitem[Schulman \& Bregman(1995)]{schulman} Schulman, E.~\& Bregman, J.~N.\ 1995, \apj, 441, 568
\bibitem[Takano et al.(1994)]{takano} Takano, M., Mitsuda, K., Fukazawa, Y., \& Nagase, F.\ 1994, \apjl, 436, L47
\bibitem[van den Bergh(1991)]{vdb}van den Bergh, S.\ 1991, \pasp, 103, 609
\bibitem[Wilms et al.(2001)]{wilms} Wilms, J., Nowak, M.~A., Pottschmidt, K., Heindl, W.~A., Dove, J.~B., \& Begelman, M.~C.\ 2001, \mnras, 320, 327
\end{thebibliography}
\end{document}